\documentclass[12pt]{article}
\usepackage{amssymb}
\usepackage{amsmath}
\def\be{\begin{equation}}
\def\ee{\end{equation}}
\def\arr{\begin{array}{rll}}
\def\ea{\end{array}}
\def\bea{\begin{eqnarray}}
\def\eea{\end{eqnarray}}

\def\N2{$N{=}2$}

\def\>{\rangle}
\def\<{\langle}
\def\+{\dagger}
\def\={\ =\ }

\begin{document}
\renewcommand{\thefootnote}{\arabic{footnote}}
\begin{titlepage}
\setcounter{page}{0}
\begin{flushright}
$\quad$  \\
\end{flushright}
\vskip 0.5cm
\begin{center}
{\LARGE\bf On the Schwarzian counterparts}\\
\vskip 0.5cm
{\LARGE\bf of conformal mechanics}\\
\vskip 2cm
$
\textrm{\Large Sergei Filyukov\ }^{a}, ~ \textrm{\Large Ivan Masterov\ }^{a,b}
$
\vskip 0.7cm
${}^{a}$
{\it
Tomsk State University of Control Systems and Radioelectronics,\\
634050, Tomsk, Lenin Ave. 40, Russia} \\
\vskip 0.5cm
${}^{b}$
{\it
Tomsk Polytechnic University, 634050, Tomsk, Lenin Ave. 30, Russia} \\
\vskip 1cm
{E-mails: filserge@tpu.ru, ivan.v.masterov@tusur.ru}

\end{center}
\vskip 1cm
\begin{abstract}
\noindent
It is shown that, if the energy in the Schwarzian mechanics (SM) is equal to the coupling constant in the de Alfaro-Fubini-Furlan (DAFF) model, there exists a link between these two systems. In particular, the equation of motion, $SL(2,\mathbb{R})$-symmetry transformations and the corresponding conserved charges of SM can be derived from those of the DAFF model by applying a coordinate transformation of a special type, while the general solution of the DAFF system maps to the velocity function of SM. A way to reproduce this link via the method of nonlinear realizations is presented. Schwarzian counterparts of the DAFF mechanics in Newton-Hooke (NH) spacetime as well as a higher derivative generalization of the DAFF model are discussed.

\end{abstract}

\vskip 1cm
\noindent
PACS numbers: 11.30.-j, 11.25.Hf, 02.20.Sv

\vskip 0.5cm

\noindent
Keywords: conformal mechanics, Schwarzian derivative, Pais-Uhlenbeck oscillator

\end{titlepage}

\noindent
{\large\bf 1. Introduction}
\vskip 0.5cm

For an arbitrary function $\rho = \rho(t)$ of one variable $t$, the following expression\footnote{Here and in what follows a number of dots over a function of one variable as well as the upper superscript in braces, which is attached to the function, designate the number of derivatives with respect to this variable.}
\bea\label{sd}
&&
\big\{\rho(t),t\big\} = \frac{\dddot{\rho}(t)}{\dot{\rho}(t)} - \frac{3}{2}\left(\frac{\ddot{\rho}(t)}{\dot{\rho}(t)}\right)^2
\eea
defines the so-called Schwarzian derivative \cite{Lagrange}-\cite{Ovsienko_1}. Originally, this object appeared in such areas of mathematics as differential equations, theory of conformal mapping, projective differential geometry (for a review see, e.g., \cite{Osgood,Ovsienko_2}).

The Schwarzian derivative holds invariance under the linear-fractional transformations
\bea\label{mob_tr}
&&
\rho'(t) = \frac{a\rho(t) + b}{c\rho(t) + d}, \qquad \det{A} = ad - cb \neq 0.
\eea
For real parameters $a$, $b$, $c$, and $d$, these transformations form an $SL^{\pm}(2,\mathbb{R})/\mathbb{Z}_{2}$ group\footnote{It is so because two sets of parameters $$(a,b,c,d)\mbox{ and }\left(\pm\frac{a}{\sqrt{|\det{A}|}},\pm\frac{b}{\sqrt{|\det{A}|}},\pm\frac{c}{\sqrt{|\det{A}|}},\pm\frac{d}{\sqrt{|\det{A}|}}\right)$$ correspond to one and the same transformation \eqref{mob_tr}.}. This property allows one to construct $SL(2,\mathbb{R})$-invariant model by using the Schwarzian derivative \eqref{sd} as a Lagrangian:
\bea\label{SM}
&&
S = \frac{1}{2} \int dt \big\{\rho(t),t\big\}.
\eea
In the literature, the model \eqref{SM} is referred to as the Schwarzian mechanics (SM) (see, e.g., \cite{Verlinde}-\cite{AV_2}). Recently, the mechanics of such type has attracted considerable interest due to its relations to the low-energy limit of the so-called Sachdev-Ye-Kitaev model \cite{Maldacena}.

As it is well-known, the so-called de Alfaro-Fubini-Furlan (DAFF) mechanics \cite{AFF,GCM}, which is described by the action functional
\bea\label{AFF}
&&
S = \int dt \, \phi(t) \left(\ddot{\phi}(t) + \frac{g^2}{\phi^3(t)}\right),
\eea
also reveals $SL(2,\mathbb{R})$-invariance. As it was noted in works \cite{Cadoni}-\cite{BS_1}, the action \eqref{AFF} takes the form
\bea\label{AFFtoSM}
&&
S = \int dt\,\left(\phi(t)\left(\ddot{\phi}(t) + \frac{g^2}{\phi^3(t)}\right) + \frac{1}{2}\big\{\rho(t),t\big\}\phi^2(t)\right)
\eea
after applying the following change of temporal and dynamical variables
\bea\label{change}
&&
t \rightarrow \rho(t), \qquad \phi(t)\rightarrow \sqrt{\dot{\rho}(t)}\cdot \phi(t).
\eea
If we formally put $\phi(t) = 1$ in \eqref{AFFtoSM}\footnote{We may put $\phi(t) = c$, where $c$ is an arbitrary constant. Then \eqref{AFFtoSM} takes the form $S = \frac{c^2}{2}\int dt \big\{\rho(t),t\big\}$. When the low energy limit of the SYK model is considered, $c^{2}$ place the role of a coupling constant in the theory (see, e.g., \cite{Verlinde}). We thank A. Galajinsky for pointing this out to us.}, we arrive to the action of SM \eqref{SM}.
\bea\label{Change}
&&
t \rightarrow \rho(t), \qquad \phi(t)\rightarrow \sqrt{\dot{\rho}(t)}.
\eea
connects actions of SM \eqref{SM} and the DAFF model \eqref{AFF}. This poses several natural questions:
\vskip 0.5cm
\noindent
1. Can such attributes of SM as the equation of motion, $SL(2,\mathbb{R})$-symmetry transforma\-tions and the corresponding conserved charges be derived from those of the DAFF model by applying the same coordinate transformation \eqref{Change}?
\vskip 0.5cm
\noindent
2. A dimensionful coupling constant $g^{2}$ enters the DAFF mechanics \eqref{AFF}. On the other hand, the action functional of SM \eqref{SM} does not contain any dimensionful constants. What is the role of the constant $g^2$ for SM with respect to the transformation \eqref{Change}?
\vskip 0.5cm
\noindent
3. Orders of differential equations that govern SM \eqref{SM} and the DAFF model \eqref{AFF} are different. Are there any relations between the trajectories of these systems under the change of coordinates {\eqref{Change}}?
\vskip 0.5cm
\noindent
4. As it is known, the DAFF model and SM can be obtained via the method of nonlinear realizations \cite{AV_2,GCM,MNR} (see also \cite{AV_6}). Can the transformation \eqref{Change} be reproduced by using the same method?
\vskip 0.5cm
\noindent
5. If a function $\rho(t)$ in \eqref{Change} satisfies the equation
\bea\label{NHrho}
&&
\big\{\rho(t),t\big\} = -2\Lambda, \qquad \Lambda = \pm\frac{1}{R^2},
\eea
the action \eqref{AFFtoSM} reproduces the analogue of the DAFF mechanics in Newton-Hooke (NH) spacetime with positive (upper sign) or negative (lower sign) cosmological constant $\Lambda$ \cite{BLL,GP}. In this interpreta\-tion, $R$ is called the characteristic time. On the other hand, the well-known Niederer's transformation, which has the form \cite{Niederer}
\bea\label{Lambda}
&&
\begin{aligned}
&
(a) &&\Lambda<0:\; t\rightarrow R\tan\frac{t}{R}, && \phi \rightarrow \frac{\phi}{\cos{t/R}},
\\[2pt]
&
(b) && \Lambda>0:\; t\rightarrow R\tanh\frac{t}{R}, && \phi \rightarrow \frac{\phi}{\cosh{t/R}},
\end{aligned}
\eea
also maps the DAFF mechanics to its NH analogue. Taking into account that the general solution of the equation \eqref{NHrho} has three arbitrary constants of integration, it is natural to ask: does the transformation \eqref{change} provide a more general construction than Niederer's transformation \eqref{Lambda}?
\vskip 0.5cm
\noindent
6. Do Schwarzain counterparts of other conformally invariant systems exist? For example, the DAFF mechanics in NH spacetime mentioned above also exhibits conformal invariance. What happens if we apply the coordinate transformation \eqref{Change} to this system? In the work \cite{Baranovsky}, a higher derivative generalization of the DAFF model has been constructed. Can the transformation \eqref{Change} be generalized to the case of higher derivative DAFF mechanics?
\vskip 0.5cm

The purpose of the present work is to investigate relationships between SM and the DAFF model and to answer the questions listed above.

The paper is organized as follows. In the next section, we show that the correspondence between SM and the DAFF model with respect to \eqref{Change} takes place only if the energy of SM is equal to the coupling constant of the DAFF model. We establish that in this case, the transformation \eqref{Change} links the equations of motion as well as the conserved charges. While the general solution of the DAFF mechanics transforms into a velocity function of SM.

In Sect. 3, we show that some particular solutions of the equation \eqref{NHrho}, when transformed under \eqref{Change}, reproduce Niederer's transformation \eqref{Lambda}. We establish that the desire to have an identical transformation in the limit $R\rightarrow\infty$ and the invariance of the DAFF model and its NH analogue under time translations unambiguously fix these particular solutions.

In Sect. 4, the equation of motion as well as a geometric description of SM associated with the DAFF model in NH spacetime are derived via the method of nonlinear realizations. The same method is also applied to reproduce the transformation \eqref{Change}.

In Sect. 5, we consider a higher derivative generalization of the DAFF mechanics. We modify transformation \eqref{Change} so as to obtain the Schwarzian counterpart for this model. In the concluding Sect. 6 we summarize our results and discuss further possible developments.

\vskip 1cm
\noindent
{\large\bf 2. Relations between the DAFF model and SM}
\vskip 0.5cm
\noindent
{\bf 2.1. The Schwarzian mechanics}
\vskip 0.5cm

Let us recall some basic facts about SM \eqref{SM}. The dynamics of this system is governed by the following equation of motion
\bea\label{EOM}
&&
\frac{1}{2\dot{\rho}}\left(\frac{\rho^{(4)}}{\dot{\rho}} - \frac{4\ddot{\rho}\rho^{(3)}}{\dot{\rho}^2} + \frac{3\ddot{\rho}^3}{\dot{\rho}^3}\right) = \frac{1}{2\dot{\rho}} \frac{d}{dt}\big\{\rho(t),t\big\} = 0.
\nonumber
\eea
Integration with respect to $t$ yields
\bea\label{SM_EOM}
&&
\big\{\rho(t),t\big\} = 2\lambda,
\eea
where $\lambda$ is a constant.

The infinitesimal form of $SL(2,\mathbb{R})$ transformations \eqref{mob_tr} reads\footnote{The transformations \eqref{SM_sym} correspond to the following choice of parameters $(a,b,c,d)$ in \eqref{mob_tr}: 
translations $\rightarrow (1,\alpha_{-1},0,1)$, dilatations $\rightarrow (e^{\alpha_{0}/2},0,0,e^{-\alpha_0/2})$, special conformal transformations $\rightarrow(1,0,-\alpha_{1},1)$.}
\bea\label{SM_sym}
&&
t' = t, \qquad \rho'(t') = \rho(t) + \alpha_{-1} + \alpha_{0}\rho(t) + \alpha_{1}\rho^{2}(t),
\eea
where constants $\alpha_{-1}$, $\alpha_{0}$ and $\alpha_{1}$ are infinitesimal parameters. One more symmetry transformation of SM reflects with the invariance of the action functional \eqref{SM} under time translations:
\bea\label{time_tr}
&&
t' = t + \sigma,\qquad \rho'(t') = \rho(t).
\eea

The generators, which correspond to the transformations \eqref{SM_sym}, \eqref{time_tr}
\bea
H = i\frac{\partial}{\partial t}, \qquad L_{-1} = i\frac{\partial}{\partial\rho}, \qquad L_{0} = i\rho\frac{\partial}{\partial\rho}, \qquad L_{1} = i\rho^2\frac{\partial}{\partial\rho},
\eea
obey commutation relations of $sl(2,\mathbb{R})\oplus\mathbb{R}$ Lie algebra
\bea\label{so12}
&&
(a):\;[H,L_{n}] = 0, \qquad (b):\;[L_{n},L_{m}] = i(m-n)L_{n+m}.
\eea

Taking into account Eq. \eqref{SM_EOM}, the conserved charges\footnote{If the action functional $S = \int dt\, L(x_i(t),\dot{x}_i(t),...,x_i^{(N)}(t))$ holds invariant under transformations of the form $t' = t + \delta t,\; x_i'(t') = x_i(t) + \delta x_i$ up to a total time derivative of some function $F = F(t)$, i.e. $\delta S = \int dt\frac{dF}{dt}$, then the corresponding integral of motion can be derived by using the expression $$L\delta t + \sum\limits_{n=0}^{N-1}\frac{d^{n}}{dt^{n}}(\delta x_i - \dot{x}_i\delta t)\sum\limits_{k = 0}^{N - n - 1}(-1)^{k}\frac{d^{k}}{dt^k}\frac{\partial L}{\partial x_i^{n+k+1}} - F.$$} associated with symmetry transforma\-tions \eqref{SM_sym} can be written as\footnote{About some systems whose Hamiltonians coincide with the $sl(2,\mathbb{R})$-Casimir operator see, e.g., \cite{MP_3}.}
\bea\label{SM_CC}
\begin{aligned}
&
\mathcal{L}_{-1} = \frac{1}{\dot{\rho}(t)} \left(\lambda + \left(\frac{\ddot{\rho}(t)}{2\dot{\rho}(t)}\right)^2\right), && \mathcal{L}_{0} = \rho(t) \mathcal{L}_{-1} - \frac{\ddot{\rho}(t)}{2\dot{\rho}(t)},
\\[2pt]
&
\mathcal{L}_{1} = -\rho^2(t) \mathcal{L}_{-1} + 2\rho(t) \mathcal{L}_{0} + \dot{\rho}(t), && \mathcal{H} = \mathcal{L}_{-1}\mathcal{L}_{1} - \mathcal{L}_{0}^2 = \lambda.
\end{aligned}
\eea
Here and in what follows we designate constants of the motion by the same letters as the correspon\-ding symmetry generators but in a calligraphic style.

Conserved charges \eqref{SM_CC} allow one to express $\dot{\rho}(t)$ in terms of these integrals of motion and $\rho(t)$:
\bea\label{SM_vel}
&&
\dot{\rho}(t) = \rho^2(t) \mathcal{L}_{-1} - 2\rho(t) \mathcal{L}_{0} + \mathcal{L}_{1} = \frac{(\rho(t) \mathcal{L}_{-1} - \mathcal{L}_{0})^2 + \lambda}{\mathcal{L}_{-1}}.
\eea
As a consequence, one may readily find the general solution to the equation \eqref{SM_EOM}.  For $\mathcal{L}_{-1}\neq 0$, this solution has the form
\bea\label{SM_GS}
&&
\rho(t) = \left\{
\begin{aligned}
&
\frac{\mathcal{L}_{0}}{\mathcal{L}_{-1}} - \frac{\sqrt{\lambda}}{\mathcal{L}_{-1}\tan{\left(\sqrt{\lambda}(t - C)\right)}}, && \lambda > 0,
\\[2pt]
&
\frac{\mathcal{L}_{0}}{\mathcal{L}_{-1}} - \frac{1}{\mathcal{L}_{-1}(t - C)}, && \lambda = 0,
\\[2pt]
&
\frac{\mathcal{L}_{0}}{\mathcal{L}_{-1}} - \frac{\sqrt{-\lambda}}{\mathcal{L}_{-1}\tanh{\left(\sqrt{-\lambda}(t - C)\right)}}, && \lambda < 0,
\end{aligned}
\right.
\eea
where $C$ is a constant of integration.

Let us discuss the case when the energy $\mathcal{H} = \lambda$ of the system \eqref{SM} is positive. Then according to the expression for conserved charge $\mathcal{L}_{-1}$ in \eqref{SM_CC}, all solutions of the equation \eqref{SM_EOM} can be divided into two disjoint parts. The first part is related to the positive-definite velocity functions $\dot{\rho}(t)$ (for negative $\mathcal{L}_{-1}$) while the second one corresponds to the negative-definite $\dot{\rho}(t)$ (for negative $\mathcal{L}_{-1}$). Each member of any part can be mapped into some instance of another by reflection
\bea\label{SM_refl}
&&
\rho(t) \rightarrow -\rho(t).
\eea
For this discrete symmetry of SM, the $SL(2,\mathbb{R})$-conserved charges in \eqref{SM_CC} are transformed as follows
\bea\label{SM_CCrefl}
&&
\mathcal{L}_{n}\rightarrow (-1)^{n}\mathcal{L}_{n}.
\eea
Analogous change of $sl(2,\mathbb{R})$-generators does not change the structure relations of the algebra \eqref{so12}, i.e. it is an automorphism of this algebra.

It is natural to suggest that the coordinate transformation \eqref{Change} is appropriate for dealing with solutions of SM which correspond to positive-definite velocity functions. While solutions with $\dot{\rho}(t)<0$ can be treated with the aid of the change
\bea\label{Change_2}
&&
t \rightarrow -\rho(t), \qquad \phi(t)\rightarrow \sqrt{-\dot{\rho}(t)},
\eea
which can be obtained from \eqref{Change} by applying \eqref{SM_refl}.

\vskip 1cm
\noindent
{\bf 2.2. A link between the DAFF model and SM}
\vskip 0.5cm

As is known \cite{AFF}, $sl(2,\mathbb{R})$-symmetry transformations, which leave invariant the DAFF mechanics \eqref{AFF}, have the form
\bea\label{AFFtransf}
&&
\mathrm{(a)}:\;t' = t + \alpha_{-1} + \alpha_{0}t + \alpha_{1}t^2, \qquad \mathrm{(b)}:\; \phi'(t') = \phi(t) + \frac{\alpha_{0}}{2}\phi(t) + \alpha_{1}t\,\phi(t).
\eea
For these symmetries, the Noether theorem yields the following constants of the motion
\bea\label{AFF_CC}
&&
\mathcal{L}_{-1} = \dot{\phi}^2(t) + \frac{g^2}{\phi^2(t)}, \quad \mathcal{L}_{0} = t \mathcal{L}_{-1} - \phi(t)\dot{\phi}(t), \quad \mathcal{L}_{1} = -t^2 \mathcal{L}_{-1} + 2t \mathcal{L}_{0} + \phi^2(t).
\eea
The integral of motion associated with the Casimir operator of $sl(2,\mathbb{R})$ is equal to the coupling constant $g^2$:
\bea\label{AFF_Cas}
&&
\mathcal{H} = \mathcal{L}_{-1}\mathcal{L}_{1} - \mathcal{L}_{0}^2 = g^2.
\eea
So, the conserved charges \eqref{AFF_CC} are functionally dependent.

The equation of motion of the DAFF mechanics \eqref{AFF}
\bea\label{AFF_EOM}
&&
\phi^3(t)\ddot{\phi}(t) = g^2
\eea
can be solved by using the integrals of motion \eqref{AFF_CC}. Indeed, by using the identity
\bea
&&
\phi^{2}(t) = t^2 \mathcal{L}_{-1} - 2 t \mathcal{L}_{0} + \mathcal{L}_{1}
\nonumber
\eea
and by taking into account the relation \eqref{AFF_Cas}, one readily obtains
\bea\label{AFF_GS}
&&
\phi^2(t) = \frac{(t \mathcal{L}_{-1} - \mathcal{L}_{0})^{2} + g^2}{\mathcal{L}_{-1}}.
\eea

Let us discuss, in which way the symmetry transformations \eqref{AFFtransf}, the equation of motion \eqref{AFF_EOM}, the conserved charges \eqref{AFF_CC} and the solution \eqref{AFF_GS} are transformed under \eqref{Change} for $\dot{\rho}(t)>0$ (or \eqref{Change_2}). Firstly, it can be verified that (\ref{AFFtransf}a) and (\ref{AFFtransf}b) are mapped into \eqref{SM_sym}
\bea
&&
\dot{\rho}'(t) = \dot{\rho}(t) + \alpha_{0}\dot{\rho}(t) + 2\alpha_1\rho(t)\dot{\rho}(t),
\nonumber
\eea
respectively. The latter is a consequence of \eqref{SM_sym}.

Secondly, the equation of motion of the DAFF model \eqref{AFF_EOM} is linked to the equation of SM \eqref{SM_EOM} only if $\lambda = g^{2}$. This means that the correspondence between equations of motion exists only if the energy of SM is equal to the coupling constant in the DAFF model. The same restriction must be satisfied so that the conserved charges of SM \eqref{SM_CC} can be derived from those of the DAFF model \eqref{AFF_CC}.

Thirdly, the general solution of the DAFF model \eqref{AFF_GS} is mapped to the expression for the velocity function \eqref{SM_vel} of SM. So, when the correspondence \eqref{Change}, \eqref{Change_2} is considered, constants $\mathcal{L}_{-1}$, $\mathcal{L}_{0}$ and $g^{2}$ of the DAFF system fix three constants of integration in the general solution of SM \eqref{SM_GS}. While the fourth constant $C$ has no DAFF prototype. At least, for this reason, the coordinate transformations \eqref{Change}, \eqref{Change_2} do not provide a one-to-one correspondence between the DAFF model and SM: one DAFF trajectory corresponds to infinitely many trajectories in SM.

\vskip 1cm
\noindent
{\large\bf 3. DAFF mechanics in NH spacetime and its Schwarzian counterpart}
\vskip 0.5cm
\noindent
{\bf 3.1. Niederer's transformation for the DAFF model from SM}
\vskip 0.5cm

Let us consider the action functional
\bea\label{AFF_NH}
&&
S = \int dt\,\phi(t)\left(\ddot{\phi}(t) + \frac{g^2}{\phi^3(t)} - \Lambda \phi(t)\right),
\eea
where $\Lambda$ is defined in \eqref{NHrho}. This action functional describes the analogue of the DAFF model \eqref{AFF} in Newton-Hooke (NH) space-time with a universal cosmological attraction ($\Lambda < 0$) or repulsion ($\Lambda > 0$) \cite{BLL,GP}. As is well known, the action \eqref{AFF_NH} can be derived from \eqref{AFF} by applying the so-called Niederer transformation \cite{Niederer}\footnote{About another transformations which connect the systems \eqref{AFF} and \eqref{AFF_NH} see also \cite{AV_7,MP_1}}. For the case of a negative cosmological constant, this transformation is given by (\ref{Lambda}a)\footnote{When $g^2=0$, the transformation (\ref{Lambda}a) relates the motion of a free particle to a half-period of harmonic oscillator.}.

As was mentioned in the Introduction, the action functional \eqref{AFFtoSM} coincides with \eqref{AFF_NH} if the function $\rho(t)$ in \eqref{AFFtoSM} satisfies the equation \eqref{NHrho}. According to \eqref{SM_GS}, this restriction can be met if
\bea\label{nied_1}
&&
\rho(t) = \frac{\mathcal{L}_{0}}{\mathcal{L}_{-1}} - \frac{1}{R\mathcal{L}_{-1}}\cdot\cot{\frac{(t - C)}{{R}}}.
\eea
This solution results in the following coordinate transformations
\bea\label{concl}
&&
t\rightarrow\pm \frac{\mathcal{L}_{0}}{\mathcal{L}_{-1}} \mp \frac{1}{R\mathcal{L}_{-1}}\cdot\cot{\frac{(t - C)}{{R}}},\qquad \phi(t)\rightarrow \frac{\phi(t)}{\sqrt{\pm\mathcal{L}_{-1}}R\sin{(t-C)/R}},
\eea
after taking into account \eqref{Change} (upper sign) and \eqref{Change_2} (lower sign). At first sight, this transformation may be more general than Niederer's one (\ref{Lambda}a). But the presence of the constants $\mathcal{L}_{0}/\mathcal{L}_{-1}$ and $C$ can be associated with the invariance of the DAFF model \eqref{AFF} and its NH analogue \eqref{AFF_NH} under time translations. Because of this, we may put
\bea
&&
\mathcal{L}_{0} = 0, \; C = -\frac{\pi R}{2}\quad \Rightarrow\quad t\rightarrow \pm\frac{1}{R\mathcal{L}_{-1}}\tan{\frac{t}{{R}}},\; \phi(t)\rightarrow \frac{\phi(t)}{\sqrt{\pm\mathcal{L}_{-1}}R\cos{t/R}},
\nonumber
\eea
without loss generality, while the constant $\mathcal{L}_{-1}$ is fixed if we require that the transformation is identical in the limit $R\rightarrow\infty$:
\bea
&&
\mathcal{L}_{-1} = \pm\frac{1}{R^2}.
\nonumber
\eea
Then we arrive to the Niederer's transformation (\ref{Lambda}a).

Relations between the general solution of SM \eqref{SM_GS} and Niederer's transformation for $\Lambda>0$ (\ref{Lambda}b) can be established from the analysis above for $\Lambda<0$ by implementing the formal change of characteristic time 
\bea\label{RiR}
&&
R\rightarrow iR.
\eea
In what follows a consideration of the DAFF mechanics in NH spacetime will be restricted by the discussion of the case of a negative cosmological constant only. While the case of a positive cosmological constant can be treated by implementing the same change of the characteristic time \eqref{RiR}.

\newpage
\noindent
{\bf 3.2. The DAFF mechanics in NH spacetime}
\vskip 0.5cm

Let us consider the model \eqref{AFF_NH} for the case of a negative cosmological constant \cite{AFF}
\bea\label{AFFNH-}
&&
S = \int dt\,\phi(t)\left(\ddot{\phi}(t) + \frac{g^2}{\phi^3(t)} + \frac{\phi(t)}{R^2}\right).
\eea
The equation of motion of this system is given by
\bea\label{AFFinHT_EOM}
&&
\ddot{\phi}(t) = \frac{g^2}{\phi^3(t)} - \frac{\phi(t)}{R^2}.
\eea

The model \eqref{AFFNH-} holds invariance under symmetry transformations
\bea\label{AFFinHT_tr}
&&
t' = t + \alpha_{-1} + \frac{R}{2}\sin{\frac{2t}{R}}\alpha_{0} + R^2\sin^2{\frac{t}{R}}\alpha_{1},
\\[2pt]
&&
\phi'(t') = \phi(t) + \frac{1}{2}\cos{\frac{2t}{R}}\phi(t)\alpha_{0} + \frac{R}{2}\sin{\frac{2t}{R}}\phi(t)\alpha_{1},
\eea
whose generators read
\bea\label{AFF_gen}
L_{-1} = i\frac{\partial}{\partial t}, \quad L_{0} = \frac{iR}{2}\sin{\frac{2t}{R}}\frac{\partial}{\partial t} + \frac{i}{2}\cos{\frac{2t}{R}}\phi\frac{\partial}{\partial\phi}, \quad L_{1} = iR^2\sin^{2}{\frac{t}{R}}\frac{\partial}{\partial t} + \frac{iR}{2}\sin{\frac{2t}{R}}\phi \frac{\partial}{\partial\phi}.
\eea
These generators form $sl(2,\mathbb{R})$ Lie algebra with commutation relations
\bea\label{so12NH}
&&
[L_{-1},L_{0}] = i\left(L_{-1} - \frac{2}{R^2}L_{1}\right), \quad [L_{-1},L_{1}] = 2iL_{0}, \quad [L_{0},L_{1}] = iL_{1}.
\eea
It is straightforward to verify that these relations can be obtained from (\ref{so12NH}b) by implementing the linear change of the basis
\bea\label{linch}
&&
L_{-1} \rightarrow L_{-1} + \frac{1}{R^2}L_{1}.
\eea

The Noether theorem yields the following conserved charges associated with $sl(2,\mathbb{R})$ symmetry transformations \eqref{AFFinHT_tr}
\bea\label{AFFinHT_CC}
&&
\mathcal{L}_{-1} = \dot{\phi}^2(t) + \frac{g^2}{\phi^2(t)} +\frac{\phi^2(t)}{R^2},\quad \mathcal{L}_{0} = \frac{R}{2}\sin{\frac{2t}{R}}\mathcal{L}_{-1} - \phi(t)\dot{\phi}(t)\cos{\frac{2t}{R}} - \frac{\phi^2(t)}{R}\sin{\frac{2t}{R}},
\nonumber
\\[2pt]
&&
\mathcal{L}_{1} = R^{2}\sin^2{\frac{t}{R}}\mathcal{L}_{-1} - \phi(t)\dot{\phi}(t)R\sin{\frac{2t}{R}} + \phi^2(t)\cos{\frac{2t}{R}}.
\eea
Taking into account the identity
\bea
&&
\phi^2(t) = \mathcal{L}_{-1} R^2\sin^2{\frac{t}{R}} - \mathcal{L}_{0} R \sin{\frac{2t}{R}} + \mathcal{L}_{1}\cos{\frac{2t}{R}}
\nonumber
\eea
and the expression for the integral of motion which corresponds to the $sl(2,\mathbb{R})$-Casimir operator
\bea
&&
\mathcal{H} = \mathcal{L}_{-1}\mathcal{L}_{1} - \mathcal{L}_{0}^{2} - \frac{\mathcal{L}_1^2}{R^2} = g^2,
\nonumber
\eea
one obtains the general solution to the equation \eqref{AFFinHT_EOM}
\bea\label{NHAFF_GS}
&&
\phi^2(t) = \frac{\left(\mathcal{L}_{1}\cos{\frac{t}{R}} - \mathcal{L}_{0}R\sin{\frac{t}{R}}\right)^2 + g^{2}R^2\sin^{2}{\frac{t}{R}}}{\mathcal{L}_{1}}.
\eea

As was mentioned above, the action functional \eqref{AFFNH-} can be obtained from the action of the DAFF mechanics \eqref{AFF} by applying Niederer's transformation (\ref{Lambda}a). By the same argument, the equation of motion \eqref{AFFinHT_EOM}, the symmetry transformations \eqref{AFFinHT_tr} and the associated integrals of motion \eqref{AFFinHT_CC} of the model \eqref{AFF_NH} can be obtained from corresponding expressions of the DAFF mechanics \eqref{AFF}.

\vskip 1cm
\noindent
{\bf 3.3. SM associated with the DAFF model in NH spacetime}
\vskip 0.5cm

Applying the coordinate transformation \eqref{Change} to the action functional \eqref{AFFNH-} gives
\bea\label{NHSM}
&&
S = \frac{1}{2}\int dt \left(\big\{\rho(t),t\big\} + \frac{2\dot{\rho}^2(t)}{R^2}\right).
\eea
The dynamics of this model is governed by the equation of motion
\bea\label{NHSM_EOM_1}
&&
\frac{1}{2\dot{\rho}} \frac{d}{dt}\left(\big\{\rho(t),t\big\} + \frac{2\dot{\rho}^2(t)}{R^2}\right) = 0.
\eea
Integration with respect to time yields
\bea\label{NHSM_EOM_2}
&&
\big\{\rho(t),t\big\} + \frac{2\dot{\rho}^2(t)}{R^2} = 2\lambda,
\eea
where $\lambda$ is a constant of integration.

As is known, $sl(2,\mathbb{R})\oplus\mathbb{R}$-symmetry transformations, which leave the model \eqref{NHSM} invariant, read
\bea\label{NHSM_tr1}
&&
t' = t + \sigma, \quad \rho'(t') = \rho(t) + \alpha_{-1} + \frac{R}{2}\sin{\frac{2\rho(t)}{R}}\alpha_{0} + R^2\sin^2{\frac{\rho(t)}{R}}\alpha_{1}.
\eea
Taking into account the equation \eqref{NHSM_EOM_2}, the corresponding integrals of motion can be written as follows:
\bea\label{NHSM_CC}
\begin{aligned}
&
\mathcal{L}_{-1} = \frac{1}{\dot{\rho}}\left(\lambda + \left(\frac{\ddot{\rho}}{2\dot{\rho}}\right)^{2} + \frac{\dot{\rho}^2}{R^2}\right), && \mathcal{L}_{0} = \frac{R}{2}\sin{\frac{2\rho}{R}}\left(\mathcal{L}_{-1} - \frac{2\dot{\rho}}{R^2}\right) - \frac{1}{2}\cos{\frac{2\rho}{R}}\cdot \frac{\ddot{\rho}}{\dot{\rho}},
\\[2pt]
&
\mathcal{H} = \mathcal{L}_{-1}\mathcal{L}_{1} - \mathcal{L}_{0}^{2} - \frac{\mathcal{L}_1^2}{R^2} = \lambda, && \mathcal{L}_{1} = R^2\sin^{2}{\frac{\rho}{R}} \left(\mathcal{L}_{-1} - \frac{2\dot{\rho}}{R^2}\right) - \frac{R}{2}\sin{\frac{2\rho}{R}}\cdot\frac{\ddot{\rho}}{\dot{\rho}} + \dot{\rho}.
\end{aligned}
\eea
These constants of the motion allow one to determine $\dot{\rho}$ by purely algebraic means
\bea\label{NHSM_vel}
&&
\dot{\rho} = \frac{\left(\mathcal{L}_{1}\cos{\frac{\rho}{R}} - \mathcal{L}_{0}R\sin{\frac{\rho}{R}}\right)^2 + \lambda R^2\sin^{2}{\frac{\rho}{R}}}{\mathcal{L}_{1}}.
\eea
Then one readily obtains the dynamics of $\rho(t)$
\bea
&&
\rho(t) = \left\{
\begin{aligned}
&
R\arctan{\frac{\mathcal{L}_{1}}{\mathcal{L}_{0}R - \sqrt{\lambda}R\tan{(\sqrt{\lambda}t - C)}}}, && \lambda > 0,
\\[2pt]
&
R\arctan{\frac{\mathcal{L}_{1}(t - C)}{R(1 + \mathcal{L}_{0}(t - C))}}, && \lambda = 0,
\\[2pt]
&
R\arctan{\frac{\mathcal{L}_{1}}{\mathcal{L}_{0}R - \sqrt{-\lambda}R\tanh{(\sqrt{-\lambda}t + C)}}}, && \lambda < 0,
\end{aligned}
\right.
\nonumber
\eea
where $C$ is a constant of integration.

As is known, the model \eqref{NHSM} can be obtained from SM \eqref{SM} by applying the coordinate transformation of the form
\bea\label{SMNT-}
\rho\rightarrow R\tan{\frac{\rho}{R}}.
\eea
In this sense, \eqref{NHSM} can be viewed as an NH analogue of \eqref{SM}. On the other hand, it can be verified that, if the energy of the system \eqref{NHSM} is equal to the coupling constant $g^{2}$ in the model \eqref{AFFNH-}, expressions \eqref{NHSM_EOM_2}, \eqref{NHSM_tr1}, \eqref{NHSM_CC}, \eqref{NHSM_vel} can be derived from those of the DAFF mechanics in the NH spacetime \eqref{AFFNH-} with the aid of the change \eqref{Change}. In this sense, the model \eqref{NHSM} can be viewed as the Schwarzian counterpart of the system \eqref{AFFNH-}.

\vskip 1cm
\noindent
{\large\bf 4. Geometric approach to SM}
\vskip 0.5cm
\noindent
{\bf 4.1. Maurer-Cartan one-forms and the equation of motion of SM}
\vskip 0.5cm

Following Ref. \cite{AV_2}, let us consider the Lie algebra $sl(2,\mathbb{R})\oplus\mathbb{R}$ which involves the generators $H$, $L_{-1}$, $L_{0}$, and $L_{1}$. The structure relations of the algebra involve (\ref{so12}a) and \eqref{so12NH}. To obtain the equation of motion of SM \eqref{NHSM}, one considers the coset space
\bea\label{coset}
&&
G(t,\rho,s,u) = e^{itH}e^{i\rho L_{-1}}e^{is L_{1}}e^{iu L_{0}},
\eea
parameterized by coordinates $t$, $\rho$, $s$, and $u$. Left multiplication by an element $G(\sigma,\alpha_{-1},\alpha_{1},\alpha_{0})$ with infinitesimal parameters $\sigma$, $\alpha_{-1}$, $\alpha_{0}$, and $\alpha_{1}$ determines the action of the group on the coset space \eqref{coset}:
\bea
&&
G(\sigma,\alpha_{-1},\alpha_{1},\alpha_{0})\cdot G(t,\rho,s,u) = G(t + \delta t, \rho + \delta\rho, s + \delta s, u + \delta u),
\nonumber
\eea
where
\bea\label{transf}
\begin{aligned}
&
\delta t = \sigma,\quad
\delta\rho = \alpha_{-1} + \frac{R}{2}\sin{\frac{2\rho}{R}}\alpha_{0} + R^2\sin^{2}{\frac{\rho}{R}}\alpha_{1},\quad
\delta u = \cos{\frac{2\rho}{R}}\alpha_{0} + R\sin{\frac{2\rho}{R}}\alpha_{1},
\\[2pt]
&
\delta s = -\left(s\cos{\frac{2\rho}{R}} + \frac{1}{R}\sin{\frac{2\rho}{R}}\right)\alpha_{0} + \left(\cos{\frac{2\rho}{R}} - sR\sin{\frac{2\rho}{R}}\right)\alpha_{1}.
\end{aligned}
\eea

Then let us construct the Maurer-Cartan one-forms:
\bea\label{MCF}
&&
G^{-1}dG = i(\omega_H H + \omega_{-1}L_{-1} + \omega_{0}L_{0} + \omega_{1}L_{1}),
\eea
where
\bea\label{MC}
&&
\omega_{H} = dt, \quad \omega_{-1} = e^{-u}d\rho, \quad \omega_{0} = du - 2sd\rho, \quad \omega_{1} = e^{u}(ds + s^{2}d\rho) + \frac{2\sinh{u}}{R^{2}}d\rho.
\eea

At the next step, let us put $\rho = \rho(t)$, $u = u(t)$, $s = s(t)$ and impose restrictions
\bea\label{constr}
&&
\omega_{n} = \nu_{n}\omega_{H},\qquad n = -1,0,1,
\eea
where $\nu_{n}$ are constants\footnote{In the literature conditions of such type are also referred to as the inverse Higgs constraints \cite{EA_1,EA_2}.}. Two constraints allow one to eliminate the fields $u=u(t)$ and $s=s(t)$ from the consideration:
\bea\label{eq1}
\begin{aligned}
&
\omega_{-1} = \nu_{-1}\omega_{H} && \Rightarrow && e^{-u} = \frac{\nu_{-1}}{\dot{\rho}};
\\[2pt]
&
\omega_{0} = \nu_{0}\omega_{H} && \Rightarrow && s = \frac{\ddot{\rho}}{2\dot{\rho}^2} - \frac{\nu_0}{2\dot{\rho}}.
\end{aligned}
\eea
The third restriction yields the equation \eqref{NHSM_EOM_2} with
\bea\label{eq3}
&&
\lambda = \nu_{-1}\nu_{1} + \frac{\nu_{-1}^{2}}{R^2} - \frac{\nu_0^2}{4}.
\eea
Finally, \eqref{transf} yields the symmetry transformations of SM \eqref{NHSM_tr1}.

\vskip 1cm
\noindent
{\bf 4.2. A geometric description of SM}
\vskip 0.5cm

Following Ref. \cite{GCM}, let us consider the matrix representation of the $sl(2,\mathbb{R})$ algebra in NH basis \eqref{so12NH}
\bea
&&
L_{-1} = \left(
\begin{aligned}
&
0 && 1
\\[2pt]
&
1/R^2 && 0
\end{aligned}
\right), \quad
L_{0} = \frac{i}{2}\left(
\begin{aligned}
&
-1 && 0
\\[2pt]
&
\;\; 0 && 1
\end{aligned}
\right), \quad
L_{1} = \left(
\begin{aligned}
&
0 && 0
\\[2pt]
&
1 && 0
\end{aligned}
\right).
\nonumber
\eea
Let us construct the $3d$ metric
\bea\label{metric}
&&
ds^{2}  = - Tr\left(L_{n}\omega_{n}\right)^2  =  2\omega_{-1}\omega_{1} - \frac{\omega_0^2}{2} + \frac{2\omega_{-1}^2}{R^2} = 2d\rho (ds + s du) - \frac{1}{2} du^2  + \frac{2}{R^2}d\rho^2,
\eea
where the Maurer-Cartan one forms $\omega_{n}$ are given in \eqref{MC}. It describes an Einsteinian spacetime because the corresponding Ricci tensor is equal to the metric.

The geodesic equations, which correspond to \eqref{metric}, read
\bea\label{GE}
\begin{aligned}
&
(\mathrm{a}):\;\ddot{\rho}(t) - \dot{\rho}(t)\dot{u}(t) = 0, \quad (\mathrm{b}):\; \ddot{u}(t) - 2\dot{\rho}(t)\dot{s}(t) -2s(t)\dot{\rho}(t)\dot{u}(t) = 0,
\\[2pt]
&
(\mathrm{c}):\; \ddot{s}(t) + 2s(t)\dot{\rho}(t)\dot{s}(t) + 2\left(s^2(t) + \frac{1}{R^2}\right)\dot{\rho}(t)\dot{u}(t) + \dot{s}(t)\dot{u}(t)  = 0,
\end{aligned}
\eea
where a proper time is denoted by $t$. Equations (\ref{GE}a) and (\ref{GE}b) allow one to express $\dot{u}(t)$ and $s(t)$ in terms of $\dot{\rho}(t)$ and $\ddot{\rho}(t)$:
\bea
&&
\dot{u}(t) = \frac{\ddot{\rho}(t)}{\dot{\rho}(t)}, \qquad s = \frac{\ddot{\rho}}{2\dot{\rho}^2} - \frac{\nu_0}{2\dot{\rho}},
\nonumber
\eea
where $\nu_{0}$ is a constant of integration, while Eq. (\ref{GE}c) leads to \eqref{NHSM_EOM_2}.

On the other hand, the equation (\ref{GE}a) after one integration yields
\bea\label{rho}
&&
\dot{\rho}(t) = \mu_{1}e^{u(t)},
\eea
where $\mu_{1}$ is an arbitrary constant. This relation enables one to rewrite (\ref{GE}b) and (\ref{GE}c) as follows:
\bea\label{equations}
\begin{aligned}
&
(\mathrm{a}):\;\ddot{u}(t) - 2\mu_1 e^{u(t)}\dot{s}(t) -2\mu_1 s e^{u(t)} \dot{u}(t) =  0,
\\[2pt]
&
(\mathrm{b}):\; \ddot{s}(t) + 2\mu_{1}e^{u(t)}s(t)\dot{s}(t) + 2\mu_1 e^{u(t)}\dot{u}(t)\left(s^2(t) + \frac{1}{R^2}\right) + \dot{s}(t)\dot{u}(t)  = 0.
\end{aligned}
\eea

After one integration, Eq. (\ref{equations}a) takes the form
\bea\label{aux}
&&
\dot{u}(t) - 2\mu_{1}e^{u(t)}s(t) = \mu_2 \;\Rightarrow\; s(t) = \frac{1}{2\mu_{1}}e^{-u(t)}\dot{u}(t) - \frac{\mu_2}{2\mu_1}e^{-u(t)},
\eea
where $\mu_2$ is also an arbitrary constant. By taking $\rho$ as an independent variable instead of $t$, one rewrites \eqref{aux} in the following way:
\bea
&&
s(\rho) = \frac{1}{2}\frac{du(\rho)}{d\rho} - \frac{\mu_2}{2\mu_1} e^{-u(\rho)}.
\nonumber
\eea
At the same time (\ref{equations}b) gives
\bea\label{auxequ}
&&
\frac{1}{2}\frac{d^{2}u(\rho)}{d\rho^{2}} + \frac{1}{4}\left(\frac{du(\rho)}{d\rho}\right)^{2} + \frac{1}{R^2} = \frac{\lambda}{\mu_{1}^2} e^{-2u(\rho)}.
\eea

Introducing the new variable
\bea\label{phi}
&&
\phi(\rho) = \sqrt{|\dot{\rho}(t)|} = \sqrt{|\mu_1|}e^{u(\rho)/2}
\eea
one can rewrite the equation \eqref{auxequ} as follows:
\bea\label{AFFNR}
&&
\frac{d^{2}\phi(\rho)}{d\rho^{2}} = \frac{\lambda}{\phi^{3}(\rho)} - \frac{\phi(\rho)}{R^2}.
\eea

The relation \eqref{phi} and the change of independent variable $t\rightarrow\rho$ reproduce the transformation \eqref{Change}.

\vskip 1cm
\noindent
{\large\bf 5. Schwarzian counterparts of higher derivative DAFF mechanics}
\vskip 0.5cm

In Ref. \cite{Baranovsky} it was shown that the model, which is described by the action functional\footnote{As it was shown in Ref. \cite{Gomis}, the model \eqref{HDAFF} with vanishing potential function reveals the so-called $l$-conformal Galilei symmetry \cite{Henkel,Negro}.}
\bea\label{HDAFF}
S = \int dt\left(\lambda_{ij}\phi_{i}\phi_{j}^{(2l+1)} + \frac{2lg^{2l+1}}{(\phi_{i}\phi_i)^{1/2l}}\right), \quad \lambda_{ij} = \left\{
\begin{aligned}
&
\delta_{ij} && i,j = 1,2,..,d, && \mbox{for half-integer }l,
\\[2pt]
&
\epsilon_{ij} && i,j = 1,2, && \mbox{for integer }l,
\end{aligned}
\right.
\eea
where $\epsilon_{ij}$ is the Levi-Civit\'{a} symbol, exhibits $sl(2,\mathbb{R})$ symmetry. The case of $l=1/2$ corresponds to the $d$-dimensional DAFF model, while the action \eqref{HDAFF} for other values of $l$ can be viewed as $(2l+1)$-order derivative generalization of the DAFF model \eqref{AFF}. The Lagrangian formulation for the odd-order DAFF mechanics is known only for $d=2$.

\vskip 0.5cm
\noindent
{\bf 5.1. Schwarzian version of one-dimensional third-order DAFF mechanics}
\vskip 0.5cm

The action functional of the third-order DAFF mechanics reads
\bea\label{AFF3}
&&
S = \int dt \left( \epsilon_{ij}\phi_i\dddot{\phi}_j + \frac{2g^{3}}{\sqrt{\phi_i \phi_i}} \right).
\eea
By analogy with \eqref{change}, let us consider the following coordinate transformation:
\bea\label{change_3}
&&
t\rightarrow\rho(t), \quad \phi_i(t) \rightarrow \dot{\rho}(t) \phi_i(t),
\eea
application of which to \eqref{AFF3} results in
\bea\label{AFF3NH}
&&
S = \int dt \left(\epsilon_{ij}\phi_i\dddot{\phi}_j + \frac{2g^{3}}{\sqrt{\phi_i \phi_i}} + 2\big\{\rho(t),t\big\}\epsilon_{ij}\phi_i\dot{\phi}_j\right).
\eea
If the function $\rho(t)$ obeys the equation \eqref{NHrho}, this action describes the third-order Pais-Uhlenbeck (PU) oscillator \cite{Lukierski,Masterov} in the presence of a potential function which preserves $sl(2,\mathbb{R})$-symmetry. It is straightforward to verify that Niederer's transformation \cite{Negro}-\cite{AV_5}
\bea
&&
t\rightarrow R\tan{\frac{t}{R}}, \quad \phi\rightarrow\frac{\phi}{\cos^{2}{t/R}}.
\nonumber
\eea
can be obtained from the solution of SM \eqref{SM_GS} in the same manner as it was made in subsection 3.1. 

If we put $\phi_i(t) = c_i$ in \eqref{AFF3NH}, where $c_i$ is a constant vector, then this action functional vanishes. Therefore one cannot obtain the Schwarzian counterpart for the two-dimensional third-order DAFF mechanics. Yet, let us consider a one-dimensional non-Lagrangian analogue of the model \eqref{AFF3}. The corresponding equation of motion reads
\bea\label{AFF3eom}
&&
\dddot{\phi}(t) = \frac{g^{3}}{\phi^{2}(t)}.
\eea
This equation takes the form
\bea\label{SM3_eom}
&&
\frac{d}{dt}\big\{\rho(t),t\big\} = g^{3},
\eea
after applying the transformation
\bea
&&
t\rightarrow\rho(t), \quad \phi(t) \rightarrow \dot{\rho}(t).
\nonumber
\eea
In this sense, \eqref{SM3_eom} can be viewed as the Schwarzian counterpart of \eqref{AFF3eom}.

\vskip 0.5cm
\noindent
{\bf 5.2. Schwarzian version of the fourth-order DAFF mechanics}
\vskip 0.5cm

Let us consider the one-dimensional fourth-order DAFF mechanics, which corresponds to the case of $l=3/2$ and $d=1$ in \eqref{HDAFF}
\bea\label{AFF4}
&&
S = \int dt\left(\phi(t)\phi^{(4)}(t) + \frac{3g^{4}}{(\phi(t))^{2/3}}\right).
\eea
This model is governed by the following equation of motion
\bea\label{eom4order}
&&
\phi^{(4)}(t) = \frac{g^{4}}{(\phi(t))^{5/3}}.
\eea

The invariance of the action functional \eqref{AFF4} under $sl(2,\mathbb{R})$-transformations
\bea\label{transf4order}
&&
t' = t + \alpha_{-1} + \alpha_{0}t + \alpha_{1}t^2, \quad \phi'(t') = \phi(t) + \frac{3}{2}\alpha_{0}\phi(t) + 3t\alpha_{1}\phi(t),
\eea
yields the following conserved charges
\bea\label{AFF4CC}
&&
\begin{aligned}
&
\mathcal{L}_{-1} = 2\dot{\phi}(t)\dddot{\phi}(t) - \ddot{\phi}^{2}(t) + \frac{3g^{4}}{\sqrt[3]{(\phi(t))^{2}}},
\\[2pt]
&
\mathcal{L}_{0} = t\mathcal{L}_{-1} - 3\phi(t)\dddot{\phi}(t) + \dot{\phi}(t)\ddot{\phi}(t),
\\[2pt]
&
\mathcal{L}_{1} = -t^2\mathcal{L}_{-1} + 2t\mathcal{L}_{0} + 6\phi(t)\ddot{\phi}(t) - 4\dot{\phi}^{2}(t).
\end{aligned}
\eea

By analogy with \eqref{Change}, let us consider the transformation
\bea\label{Change_3}
&&
t\rightarrow \rho(t), \qquad \phi(t) \rightarrow \sqrt{\dot{\rho}^{3}(t)},
\eea
where we assume that $\dot{\rho}(t)>0$. Then the action \eqref{AFF4} takes the form
\bea\label{SM4order}
&&
S = \frac{9}{4}\int dt\,\big\{\rho(t),t\big\}^2.
\eea
The dynamics of this system is described by the equation of motion
\bea\label{SM4eom}
\frac{1}{\dot{\rho}(t)}\frac{d}{dt}\left(\frac{3}{2}\frac{d^2}{dt^2}\big\{\rho(t),t\big\} + \frac{9}{4}\big\{\rho(t),t\big\}^{2}\right) = 0 \;\Rightarrow\; \frac{d^2}{dt^2}\big\{\rho(t),t\big\} + \frac{3}{2}\big\{\rho(t),t\big\}^{2} = \frac{2\lambda}{3},
\eea
where $\lambda$ is a constant of integration.

It is evident that the model \eqref{SM4order} holds invariance under $sl(2,\mathbb{R})$-transformations \eqref{SM_sym}. The same transformations are derived from \eqref{transf4order} by applying the change \eqref{Change_3}. It can also be verified that if one sets
\bea\label{lambda}
&&
\lambda = g^4,
\eea
the integrals of motion associated with $SL(2,\mathbb{R})$-symmetry in the model \eqref{SM4order} can be derived from \eqref{AFF4CC} with the aid of \eqref{Change_3}.

When the condition \eqref{lambda} is satisfied, \eqref{SM4eom} takes the form
\bea
&&
\ddot{y}(t) - 6y^{2}(t) = - \frac{g^{4}}{6},
\nonumber
\eea
where
\bea
&&
y(t) = -\frac{1}{4}\big\{\rho(t),t\big\}.
\nonumber
\eea
The order of this equation can be reduced by implementing the change
\bea
&&
\dot{y}(t) = p(y(t)) \;\Rightarrow\; p(y)\frac{dp(y)}{dy} - 6y^{2} = -\frac{g^{4}}{6}.
\nonumber
\eea
After one integration one obtains
\bea
&&
\dot{y}^{2}(t) = 4y^{3}(t) - \frac{g^{4}}{3}y(t) + C_1,
\nonumber
\eea
where $C_{1}$ is a constant of integration. The solution of this equation is the Weierstrass elliptic function $y(t) = \wp\left(t + C_2; g^{4}/3;C_{1}\right)$ (see, e.g., \cite{WF}). So, \eqref{SM4eom} leads to
\bea
&&
\big\{\rho(t),t\big\} = -4\wp\left(t + C_2; g^{4}/3;C_{1}\right),
\nonumber
\eea
integration of which is problematic.

On the other hand, any function $\rho = \rho(t)$ which satisfies
\bea
&&
\big\{\rho(t),t\big\} = \frac{2g^2}{3},
\nonumber
\eea
is a particular solution to \eqref{SM4eom}. Therefore according to \eqref{SM_GS} we have the following class of particular solutions to the equation \eqref{SM4eom}
\bea\label{SM4_ps}
&&
\rho(t) = \frac{\mathcal{A}_{0}}{\mathcal{A}_{-1}} - \frac{\tilde{g}}{\mathcal{A}_{-1}\tan{\left(\tilde{g}(t - C)\right)}},
\eea
where $\mathcal{A}_{-1}$ and $\mathcal{A}_{0}$ are arbitrary constants and $\tilde{g} = g/\sqrt{3}$. The velocity function, related to this class of solutions, reads
\bea
&&
\dot{\rho}(t) = \frac{(\rho(t) \mathcal{A}_{-1} - \mathcal{A}_{0})^2 + \tilde{g}^{2}}{\mathcal{A}_{-1}}.
\nonumber
\eea
Taking into account the coordinate transformation \eqref{Change_3}, we immediately obtain a class of particular solutions to \eqref{eom4order}
\bea
&&
\phi^{2}(t) = \left( \frac{(t \mathcal{A}_{-1} - \mathcal{A}_{0})^2 + \tilde{g}^{2}}{\mathcal{A}_{-1}}\right)^{3}.
\nonumber
\eea

Note that $\lambda$ in \eqref{Lambda} has the meaning of the energy of the system \eqref{SM4order}. So, the correspondence between \eqref{SM4order} and \eqref{AFF4} exists only if the energy of the model \eqref{SM4order} coincides with the coupling constant in \eqref{AFF4}.

\vskip 0.5cm
\noindent
{\bf 5.3. Schwarzian analogues of one-dimensional higher derivative DAFF dynamics}
\vskip 0.5cm

The equation of motion, which corresponds to one-dimensional $(2l+1)$-order derivative DAFF mechanics, has the form
\bea\label{AFFHD}
&&
\phi^{(2l+1)}(t) = \frac{g^{2l+1}}{\phi^{(l+1)/l}(t)}.
\eea
To obtain Schwarzian counterparts of this equation, let us introduce the transformation
\bea\label{changeHD}
&&
t\rightarrow \rho(t), \qquad \phi(t) \rightarrow (\dot{\rho}(t))^{l}.
\eea
The equation, which arises as the result of applying \eqref{changeHD} to \eqref{AFFHD}, can be represented in the form
\bea\label{SMHD_eom}
&&
l\cdot S_{2l+1}(\rho) = g^{2l+1},
\eea
where $S_{2l+1}$ are defined by the following recurrent relation
\bea
&&
S_{2}(\rho) = \big\{\rho(t),t\big\},
\nonumber
\\[2pt]
&&\label{HSch}
S_{2l+1}(\rho) = \frac{d}{dt}S_{2l}(\rho) + \frac{1}{4}\sum_{k=2}^{2l-1}\frac{(2l+1)!}{k!(2l-k+1)!}S_{k}(\rho)S_{2l-k+1}(\rho), \quad l\geq 1.
\eea
The formal symbol $\sum\limits_{k=2}^{1}f(k)$, which appears for $l=1$, is assumed to be equal to zero. The similar Schwarzian structures appeared in \cite{Aharonov}-\cite{Kim}.

Eq. \eqref{SMHD_eom} for $l=1$ and $l=3/2$ reproduces \eqref{SM3_eom} and \eqref{SM4eom}, respectively, while, for example, for $l=2$, $l=5/2$, and $l=3$ we have
\bea
&&
\dddot{S}_{2}(\rho) + 8 S_{2}(\rho) \dot{S}_{2}(\rho) = \frac{g^{5}}{2};
\nonumber
\\[2pt]
&&
S_{2}^{(4)}(\rho) + \frac{31}{2}S_{2}(\rho)\ddot{S}_{2}(\rho) + 13\dot{S}_{2}^{2}(\rho) + \frac{45}{4}S_{2}^{3}(\rho) = \frac{2g^{6}}{5};
\nonumber
\\[2pt]
&&
S_{2}^{(5)}(\rho) + 26 S_{2}(\rho)S_{2}^{(3)}(\rho) + 59\dot{S}_{2}(\rho)\ddot{S}_{2}(\rho) + 144 S_{2}^{2}(\rho)\dot{S}_{2}(\rho) = \frac{g^{7}}{3}.
\nonumber
\nonumber
\eea

NH counterpart of \eqref{AFFHD} can be obtained by applying Niederer's transformation
\bea
&&
\begin{aligned}
&
\prod_{k=0}^{l-1/2}\left(\frac{d^2}{dt^2}+\frac{(2k+1)^2}{R^2}\right)\phi(t) = \frac{g^{2l+1}}{\phi^{(l+1)/l}(t)}, && \mbox{for half-integer }l,
\\[2pt]
&
\prod_{k=1}^{l}\left(\frac{d^2}{dt^2}+\frac{(2k)^2}{R^2}\right)\dot{\phi}(t) = \frac{g^{2l+1}}{\phi^{(l+1)/l}(t)}, && \mbox{for integer }l.
\end{aligned}
\nonumber
\eea
These equations describe the conformally invariant $(2l+1)$-order PU oscillator \cite{AV_4}-\cite{Krivonos} in the external field which preserves conformal symmetry. By applying the transformation \eqref{changeHD}, we obtain the Schwarzian analogues which read as in \eqref{SMHD_eom} with
\bea
&&
S_{2}(\rho) = \big\{\rho(t),t\big\} + \frac{2\dot{\rho}^{2}(t)}{R^2},
\nonumber
\eea
while $S_{2l+1}$ for $l>1/2$ are given by the same recurrent relation \eqref{HSch}.

It should be mentioned that for any half-integer $l$ the equation \eqref{SMHD_eom} has a class of particular solutions of the form \eqref{SM4_ps} with
\bea
&&
\tilde{g} = \frac{g}{\sqrt[l+1/2]{(2l)!!}}.
\nonumber
\eea
As a consequence, the function
\bea
&&
\phi^{2}(t) = \left( \frac{(t \mathcal{A}_{-1} - \mathcal{A}_{0})^2 + \tilde{g}^{2}}{\mathcal{A}_{-1}}\right)^{2l}
\nonumber
\eea
represents a class of particular solutions to the corresponding DAFF equation \eqref{AFFHD}.

\vskip 1cm
\noindent
{\large\bf 6. Conclusion}
\vskip 0.5cm
To summarize, in this work a relationship between the DAFF model and SM was elucidated. It was demonstrated that:
\vskip 0.5cm
\noindent
1) the equation of motion, $SL(2,\mathbb{R})$-symmetry transforma\-tions, and corresponding conserved charges of SM can be derived from those of the DAFF model by applying the coordinate transformations \eqref{Change};
\vskip 0.5cm
\noindent
2) the correspondence between the models exists if the coupling constant $g^{2}$ in the DAFF model is equal to the energy of SM;
\vskip 0.5cm
\noindent
3) the link \eqref{Change} maps the solution of the DAFF model to the velocity function of SM. As a result, one DAFF trajectory corresponds to infinitely many solutions of SM;
\vskip 0.5cm
\noindent
4) the coordinate transformation \eqref{Change} can be reproduced within the method of nonlinear realizations;
\vskip 0.5cm
\noindent
5) the link \eqref{Change}, accompanied by the solution of SM, provides a more general construction \eqref{concl} than Niederer's transformation \eqref{Lambda}. But additional freedom in \eqref{concl} can be fixed by taking into account the invariance of the DAFF model and its NH counterpart under time translations and by imposing the requirement that the transformation is identical in the limit $R\rightarrow\infty$;
\vskip 0.5cm
\noindent
6) the Schwarzain counterpart of the DAFF mechanics in the NH spacetime can be derived by applying the same coordinate transformation \eqref{Change}. The link \eqref{Change} can be generalized to derive Schwarzian counterparts for one-dimensional higher derivative DAFF mechanics. This link allowed us to obtain a two-parametric set of particular solutions for the even-order DAFF mechanics.

\vskip 0.5cm
Turning to further developments, it would be interesting to investigate the relationships between the DAFF model and SM within the Hamiltonian formalism. Taking into account the results in \cite{MP_2}, it is of interest to construct possible Schwarzian counterparts of supersymmetric conformal mechanics. In particular, it is worth studying whether the analysis in this work can be generalized to the case of the $\mathcal{N}=1$ higher derivative mechanical systems obtained in \cite{Masterov_3}. These issues will be studied elsewhere.

\section*{Acknowledgements}
We are greatful to A. Galajinsky for the comments on the manuscript. We also thank E.A. Ivanov and M.S. Plyushchay for the correspondence. This work was supported by the Russian Science Foundation, grant No 19-11-00005.

\fontsize{10}{13}\selectfont


\begin{thebibliography}{nn}
\bibitem{Lagrange}
J.-L. Lagrange, {\it Sur la construction des cartes g\'{e}ographiques}, Nouveaux M\'{e}moires de l'Acad\'{e}mie de Berlin, 1779.
\bibitem{Cayley}
A. Cayley, {\it On the Schwarzian derivative and polyhedral functions}, Trans. Camb. Phil. Soc. {\bf 13} (1880).
\bibitem{Kummer}
E. Kummer, {\it \"{U}ber die hypergeometrische Reihe}. Crelle, {\bf 15} (1836) 39-83 and 127-172.
\bibitem{Schwarz}
H. Schwarz, {\it Gesammelte mathematische Abhandlungen}, Springer, Berlin, 1890.
\bibitem{Ovsienko_1}
V. Ovsienko, S. Tabachnikov, {\it What is... the Schwarzian derivative?}, Notices of the AMS {\bf 56} (2009) 34.
\bibitem{Osgood}
B. Osgood, {\it Old and New on the Schwarzian Derivative}. In: Duren P., Heinonen J., Osgood B., Palka B. (eds) {\it Quasiconformal Mappings and Analysis}, Springer, New York, 1998.
\bibitem{Ovsienko_2}
V. Ovsienko, S. Tabachnikov, {\it Projective differential geometry old and new, from Schwarzian derivative to the cohomology of diffeomorphism groups}, Cambridge Tracts in Mathematics, {\bf 165}, Cambridge University Press, 2005.
\bibitem{Verlinde}
T.G. Mertens, G.J. Turiaci, H.L. Verlinde, {\it Solving the Schwarzian via the conformal Bootstrap}, JHEP {\bf 1708} (2017) 136, arXiv:1705.08408[hep-th].
\bibitem{Verlinde_1}
H.T. Lam, T.G. Mertens, G.J. Turiaci, H. Verlinde, {\it Shockwave S-matrix from Schwarzian quantum mechanics}, JHEP {\bf 1811} (2018) 182, arXiv:1804.09834[hep-th].
\bibitem{AV_1}
A. Galajinsky, {\it A variant of Schwarzian mechanics}, Nucl. Phys. B {\bf 936} (2018) 661, arXiv:1809.00904[hep-th].
\bibitem{AV_2}
A. Galajinsky, {\it Schwarzian mechanics via nonlinear realizations}, Phys. Lett. B {\bf 795} (2019) 277, arXiv:1905.01935[math-ph].
\bibitem{Maldacena}
J. Maldacena, D. Stanford, {\it Remarks on the Sachdev-Ye-Kitaev model}, Phys. Rev. D {\bf 94} (2016) 106002, arXiv:1604.07818[hep-th].
\bibitem{AFF}
V. de Alfaro, S. Fubini, G. Furlan, {\it Conformal invariance in quantum mechanics}, Nuovo Cim. A {\bf 34} (1976) 569.
\bibitem{GCM}
E. Ivanov, S. Krivonos, V. Leviant, {\it Geometry of conformal mechanics}, J. Phys. A {\bf 22} (1989) 345.
\bibitem{Cadoni}
M. Cadoni, P. Carta, S. Mignemi, {\it A realization of the infinite-dimensional symmetries of the conformal mechanics}, Phys. Rev. D {\bf 62} (2000) 086002, hep-th/0004107.
\bibitem{Astorino}
M. Astorino, S. Cacciatori, D. Klemm, D. Zanon, {\it AdS(2) supergravity and superconformal quantum mechanics}, Annals Phys. {\bf 304} (2003) 128, hep-th/0212096.
\bibitem{Zyl}
H.J.R. van Zyl, {\it Constructing dualities from quantum state manifolds}, arXiv:1509.01231[hep-th].
\bibitem{Maldacena_2}
J. Maldacena, D. Stanford, Z. Yang, {\it Conformal symmetry and its breaking in two dimensional Nearly Anti-de-Sitter space}, PTEP {\bf 12} (2016) 12C104, arXiv:1606.01857[hep-th].
\bibitem{Grumiller}
D. Grumiller, R. McNees, J. Salzer, C. Valc\'{a}rcel, D. Vassilevich, {\it Menagerie of AdS$_2$ boundary conditions}, JHEP {\bf 1710} (2017) 203, arXiv:1708.08471[hep-th].
\bibitem{Lidsey}
J.E. Lidsey, {\it Inflationary cosmology, diffeomorphism group of the line and Virasoro coadjoint orbits}, arXiv:1802.09186[hep-th].
\bibitem{BS_2}
V.V. Belokurov, E. Shavgulidze, {\it Unusual view of the Schwarzian theory}, Mod. Phys. Lett. A {\bf 33} (2018) 1850221, arXiv:1806.05605[hep-th].
\bibitem{BS_1}
V.V. Belokurov, E.T. Shavgulidze, {\it Polar decomposition of the Wiener measure: Schwarzian theory versus conformal quantum mechnics}, Theor. Mat. Phys. {\bf 200} (2019) 1324, arXiv:1812.04039[hep-th].
\bibitem{MNR}
S.R. Coleman, J. Wess, B. Zumino, {\it Structure of phenomenological Lagrangians. I}, Phys. Rev. {\bf 177} (1969) 2239.
\bibitem{AV_6}
A. Galajinsky, {\it Super-Schwarzians via nonlinear realizations}, arXiv:2004.04489[hep-th].
\bibitem{BLL}
H. Bacry, J. Levy-Leblond, {\it Possible kinematics}, J. Math. Phys. {\bf 9} (1968) 1605.
\bibitem{GP}
G.W. Gibbons, C.E. Patricot, {\it Newton-Hooke space-times, Hpp waves and the cosmological constant}, Class. Quant. Grav. {\bf 20} (2003) 5225, hep-th/0308200.
\bibitem{Niederer}
U. Niederer, {\it The maximal kinematical invariance group of the harmonic oscillator}, Helv. Phys. Acta {\bf 46} (1973) 191.
\bibitem{MP_3}
F. Correa, V. Jakubsky, M.S. Plyushchay, {\it Aharonov-Bohm effect on AdS(2) and nonlinear supersymmetry of reflectionless Poschl-Teller system}, Annals Phys. {\bf 324} (2009) 1078, arXiv:0809.2854[hep-th].
\bibitem{AV_7}
A. Galajinsky, O. Lechtenfeld, K. Polovnikov, {\it Calogero models and nonlocal conformal transformations}, Phys. Lett. B {\bf 643} (2006) 221, hep-th/0607215.
\bibitem{MP_1}
L. Inzunza, M.S. Plyushchay, A. Wipf, {\it Conformal bridge between freedom and confinement}, arXiv:1912.11752[hep-th].
\bibitem{EA_1}
E.A. Ivanov, V.I. Ogievetsky, {\it The inverse Higgs phenomenon in nonlinear realizations}, Teor. Mat. Fiz. {\bf 25} (1975) 164.
\bibitem{EA_2}
E.A. Ivanov, V.I. Ogievetsky, {\it Inverse Higgs effect in nonlinear realizations}, Theor. Math. Phys. {\bf 25} (1975) 1050.
\bibitem{Baranovsky}
O. Baranovsky, {\it Higher-derivative generalization of conformal mechanics}, J. Math. Phys. {\bf 58} (2017) 082903, arXiv:1704.04880[hep-th].
\bibitem{Gomis}
J. Gomis, K. Kamimura, {\it Schrodinger equations for higher order non-relativistic particles and N-Galilean conformal symmetry}, Phys. Rev. D {\bf 85} (2012) 045023, arXiv:1109.3773[hep-th].
\bibitem{Henkel}
M. Henkel, {\it Local scale invariance and strongly anisotropic equilibrium critical systems}, Phys. Rev. Lett. {\bf 78} (1997) 1940, cond-mat/9610174.
\bibitem{Negro}
J. Negro, M.A. del Olmo, A. Rodriguez-Marco, {\it Nonrelativistic conformal groups}, J. Math. Phys. {\bf 38} (1997), 3786.
\bibitem{Lukierski}
J. Lukierski, P.C. Stichel, W.J. Zakrzewski, {\it Galilean invariant (2+1)-dimensional models with a Chern-Simons-like term and D=2 noncommutative geometry}, Annals Phys. {\bf 260} (1997) 224, hep-th/9612017.
\bibitem{Masterov}
I. Masterov, {\it The odd-order Pais-Uhlenbeck oscillator}, Nucl. Phys. B {\bf 907} (2016) 495, arXiv:1603.07727[math-ph].
\bibitem{Horvathy}
C. Duval, P. Horvathy, {\it Conformal Galilei groups, Veronese curves, and Newton-Hooke spacetimes}, J. Phys. A {\bf 44} (2011) 335203, arXiv:1104.1502[hep-th].
\bibitem{AV_5}
A. Galajinsky, I. Masterov, {\it Remarks on l-conformal extension of the Newton-Hooke algebra}, Phys. Lett. B {\bf 702} (2011) 265, arXiv:1104.5115[hep-th].
\bibitem{WF}
G. Pastras, {\it Four lectures on Weierstrass elliptic function and applications in classical and quantum mechanics}, arXiv:1706.07371[math-ph].
\bibitem{Aharonov}
D. Aharonov, {\it A necessary and sufficient condition for univalence of a meromorphic functions}, Duke Math. J. {\bf 36} (1969) 599.
\bibitem{Tamanoi}
H. Tamanoi, {\it Higher Schwarzian operators and combinatorics of the Schwarzian derivative}, Math. Annalen {\bf 305} (1996) 127.
\bibitem{Kim}
S. Kim, T. Sugawa, {\it Invariant Schwarzian derivatives of higher order}, Complex Analysis and Operator Theory {\bf 5} (2011) 659, arXiv:0911.2663[math].
\bibitem{AV_4}
A. Galajinsky, I. Masterov, {\it Dynamical realizations of l-conformal Newton-Hooke group}, Phys. Lett. B {\bf 723} (2013) 190, arXiv:1303.3419[hep-th].
\bibitem{AV_3}
K. Andrzejewski, A. Galajinsky, J. Gonera, I. Masterov, {\it Conformal Newton-Hooke symmetry of Pais-Uhlenbeck oscillator}, Nucl. Phys. B {\bf 885} (2014) 150, arXiv:1402.1297[hep-th].
\bibitem{Masterov_2}
I. Masterov, {\it Dynamical realizations of N=1 l-conformal Galilei superalgebra}, J. Math. Phys. {\bf 55} (2014) 102901, arXiv:1407.1438[hep-th].
\bibitem{Krivonos}
S. Krivonos, O. Lechtenfeld, A. Sorin, {\it Minimal realization of l-conformal Galilei algebra, Pais-Uhlenbeck oscillator and their deformation}, JHEP {\bf 1610} (2016) 078, arXiv:1607.03756[hep-th].
\bibitem{MP_2}
M.S. Plyushchay, {\it Schwarzian derivative treatment of the quantum second-order supersymmetry anomaly, and coupling-constant metamorphosis}, Annals Phys. {\bf 377} (2017) 164, arXiv:1602.02179[hep-th].
\bibitem{Masterov_3}
I. Masterov, B. Merzlikin, {\it Superfield approach to higher derivative N=1 superconformal mechanics}, JHEP {\bf 1911} (2019) 165, arXiv:1909.12574[hep-th].

\end{thebibliography}
\end{document}